\documentclass{PoS}

\title{Neutral kaon interferometry at KLOE and KLOE-2}

\ShortTitle{Neutral kaon interferometry at KLOE and KLOE-2}

\author{\speaker{Izabela Balwierz} \\on behalf of the KLOE and KLOE-2 collaborations\thanks{The KLOE collaboration:
F.~Ambrosino, A.~Antonelli, M.~Antonelli, F.~Archilli, I.~Balwierz, G.~Bencivenni, C.~Bini, C.~Bloise, S.~Bocchetta, F.~Bossi, P.~Branchini, G.~Capon, T.~Capussela, F.~Ceradini, P.~Ciambrone, E.~Czerwi\'nski, E.~De~Lucia, A.~De~Santis, P.~De~Simone, G.~De~Zorzi, A.~Denig, A.~Di~Domenico, C.~Di~Donato, B.~Di~Micco, M.~Dreucci, G.~Felici, S.~Fiore, P.~Franzini, C.~Gatti, P.~Gauzzi, S.~Giovannella, E.~Graziani, M.~Jacewicz, J.~Lee-Franzini, M.~Martemianov, M.~Martini, P.~Massarotti, S.~Meola, S.~Miscetti, G.~Morello, M.~Moulson, S.~M\"uller, M.~Napolitano, F.~Nguyen, M.~Palutan, A.~Passeri, V.~Patera, I.~Prado~Longhi, P.~Santangelo, B.~Sciascia, M.~Silarski, T.~Spadaro, C.~Taccini, L.~Tortora, G.~Venanzoni, R.Versaci, G.~Xu, J.~Zdebik, and, as members of the KLOE-2 collaboration: D.~Babusci, D.~Badoni, V.~Bocci, A.~Budano, S.~A.~Bulychjev, L.~Caldeira~Balkest\aa hl, P.~Campana, E.~Dan\'e, G.~De~Robertis, D.~Domenici, O.~Erriquez, G.~Fanizzi, G.~Giardina, F.~Gonnella, F.~Happacher, B.~H\"oistad, L.~Iafolla, E.~Iarocci, T.~Johansson, A.~Kowalewska, V.~Kulikov, A.~Kupsc, F.~Loddo, G.~Mandaglio, M.~Mascolo, M.~Matsyuk, R.~Messi, D.~Moricciani, P.~Moskal, A.~Ranieri, C.~F.~Redmer, I.~Sarra, M.~Schioppa, A.~Sciubba, W.~Wi\'slicki, M.~Wolke
}\\ 
        Jagiellonian University, Institute of Physics\\
        E-mail: \email{iza.balwierz@uj.edu.pl}}

\abstract{Neutral kaons produced in correlated pairs at a phi-factory offer unique possibilities to perform fundamental tests of CPT invariance, as well as of the basic principles of quantum mechanics. The analysis of the data collected by the KLOE experiment at DAFNE is still ongoing with the aim of improving previous results and limits on several parameters describing CPT violation and decoherence. Ancillary measurements like the regeneration cross section on the beam pipe materials are also in progress and will be very useful to reduce the systematic uncertainties. Prospects on improvements at the KLOE-2 experiment, aiming at an increase of the integrated luminosity of about a factor of ten with an upgraded detector, will be also discussed.}

\FullConference{8th International Conference on Nuclear Physics at Storage Rings - Stori11,\\
		October 9-14, 2011\\
		INFN, Laboratori Nazionali di Frascati, Italy}

\begin{document}

\section{The Frascati $\phi$-factory facility}

\subsection{The KLOE experiment at the DAFNE collider}

\begin{figure}[!h]
\hspace{2.cm}
\parbox{0.30\textwidth}{\centerline{\includegraphics[width=8.2cm]{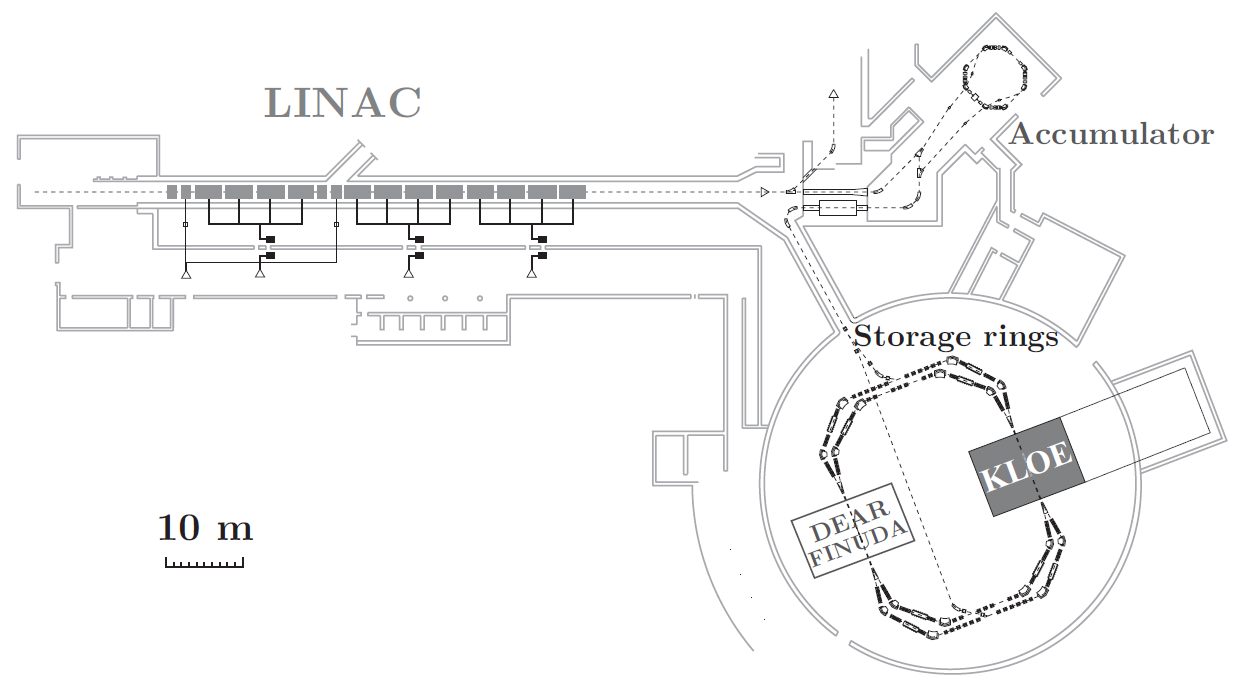}}}
\hspace{3.cm}
\parbox{0.30\textwidth}{\centerline{\includegraphics[width=6cm]{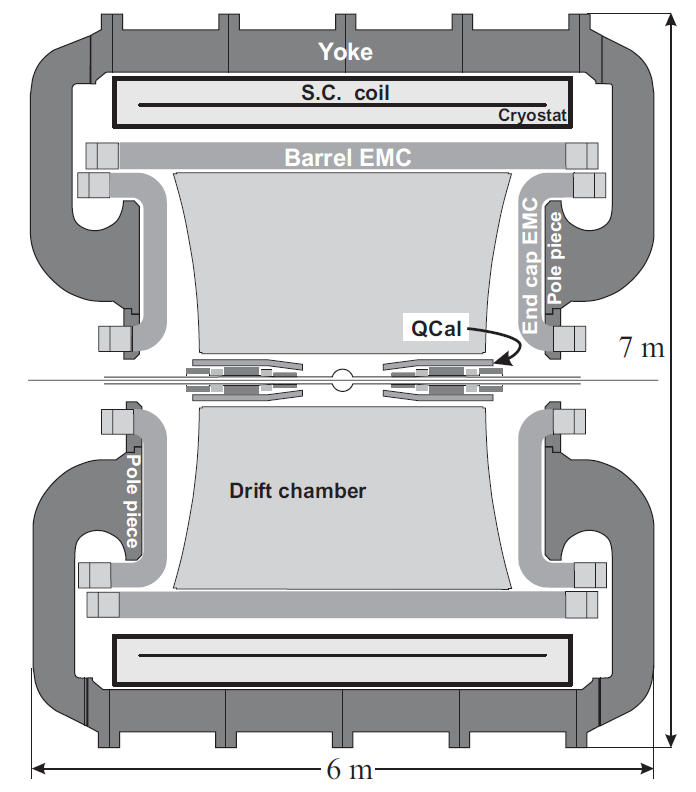}}}
\caption{Left: The DAFNE facility scheme, right: scheme of the KLOE detector. The figures are adapted from \cite{the_physics}.}
\label{dafne}
\end{figure}

The DAFNE $\phi$-factory \cite{flavour}, located at the Frascati laboratory (LNF) of INFN, is an $e^+ e^-$ accelerator, working at the $\phi$ resonance peak $\sqrt{s}=m_{\phi}\approx1019$~GeV with the $\phi$ production cross section equal to $\sigma(e^+ e^-\to\phi)\approx3.1$~$\mu$b. DAFNE consists of 3 main parts: a linear accelerator, an accumulator and two storage rings in which electrons and positrons collide (Fig. \ref{dafne} left). The KLOE detector is placed at the center of one of the two interaction points. Data were collected from 2001 to 2006 corresponding to $\sim2.5~\textrm{fb}^{-1}$ of integrated luminosity, i.e. $\sim2.5$ billion $\phi$ meson decays into neutral kaon pairs. The year 2006 was spent on collecting about 250~pb$^{-1}$ of off-peak data.

The KLOE detector (Fig. \ref{dafne} right) consists of a cylindrical drift chamber (DC) \cite{DC}, surrounded by an a electromagnetic calorimeter (EmC) \cite{EmC}, both inserted in a superconducting coil which produces an axial magnetic field of 0.52~T, parallel to the beam axis. The diameter and length of DC are 4~m and 3.3~m, respectively. The chamber is filled with 90\% of helium and 10\% of isobutan. Momentum reconstruction from the curvature of the track has a fractional accuracy of $\frac{\sigma_p}{p} \simeq 0.5\%$ . The spatial resolution is below 200~$\mu$m in the transverse plane (''$\varphi$-coordinates'') and the accuracy of vertex reconstruction is about $\simeq 1$~mm. EmC is a sampling calorimeter consisting of an altemating stack of 1~mm scintillating fiber layers glued between thin grooved lead foils. The whole calorimeter consists of a ''barrel'' and two ''end caps'' and covers almost $4\pi$ solid angle. Energy and time resolutions of this calorimeter, for a photon's energy in a range from 20 to 500~MeV, are equal to $\sigma(E)=\frac{5.7\%}{\sqrt{E(\mathrm{GeV})}}$ and $\sigma(t)=\frac{54\ \mathrm{ps}}{\sqrt{E(\mathrm{GeV})}}\oplus 50\ \mathrm{ps}$, respectively. The beam pipe of KLOE at the interaction point has a spherical shape, 500~$\mu$m thick, with 10~cm radius and is built from an alloy of beryllium and aluminum. Inside the sphere, a 50 micron thick beryllium cylinder, coaxial with the beam, provides electrical continuity.

\subsection{Neutral kaons at a $\phi$-factory}

At KLOE neutral kaons are produced in an entangled states, in a decay of the $\phi$ meson, which is a vector meson with $J^{PC}=1^{--}$. The branching ratio for the $\phi$ decay into $K^0 \bar{K}^0$ is about 34\% and at DAFNE it corresponds to about $10^6$ neutral kaons produced per pb$^{-1}$ in an antisymmetric quantum state with quantum numbers 1$^{--}$ \cite{neutral_kaon}:
\begin{eqnarray}
\left|i\right>&=&\frac{1}{\sqrt{2}}\left[\left|K^0(+\vec{p})\right>\left|\bar{K}^0(-\vec{p})\right>-\left|\bar{K}^0(+\vec{p})\right>\left|K^0(-\vec{p})\right>\right]=\nonumber\\
&=&\frac{N}{\sqrt{2}}\left[\left|K_S(+\vec{p})\right>\left|K_L(-\vec{p})\right>-\left|K_L(+\vec{p})\right>\left|K_S(-\vec{p})\right>\right],
\label{eq:initial_2}
\end{eqnarray}
where $N$ \cite{neutral_kaon} is a normalization factor. In the second row the strangeness basis $\{\left|K^0\right>$, $\left|\bar{K}^0\right>\}$, suitable to describe kaons production, is changed to $\{\left|K_S\right>$, $\left|K_L\right>\}$ basis, appropriate to describe decays of kaons.

At KLOE kaons have momenta of about 110~MeV. Due to the large lifetime difference of both kaons ($\tau_L\approx51$~ns, $\tau_S\approx90$~ps) there is also large difference in their mean decay lengths, namely for $K_S$ it is about 6~mm whereas for $K_L$ about 3.5~m. This fact enables to identify the $K_L$ mesons decays by the presence of $K_S$ decays into two pions close to the interaction region. This type of selection at KLOE is called the $K_L\ tagging$. What is unique at a $\phi$-factory and not possible at fixed target experiments is tagging of the $K_S$ mesons, which in the KLOE detector is realized by direct hits of $K_L$ into the calorimeter.

\subsection{Neutral kaon interferometry}

\begin{figure}[!h]
\centering
\includegraphics[width=12cm]{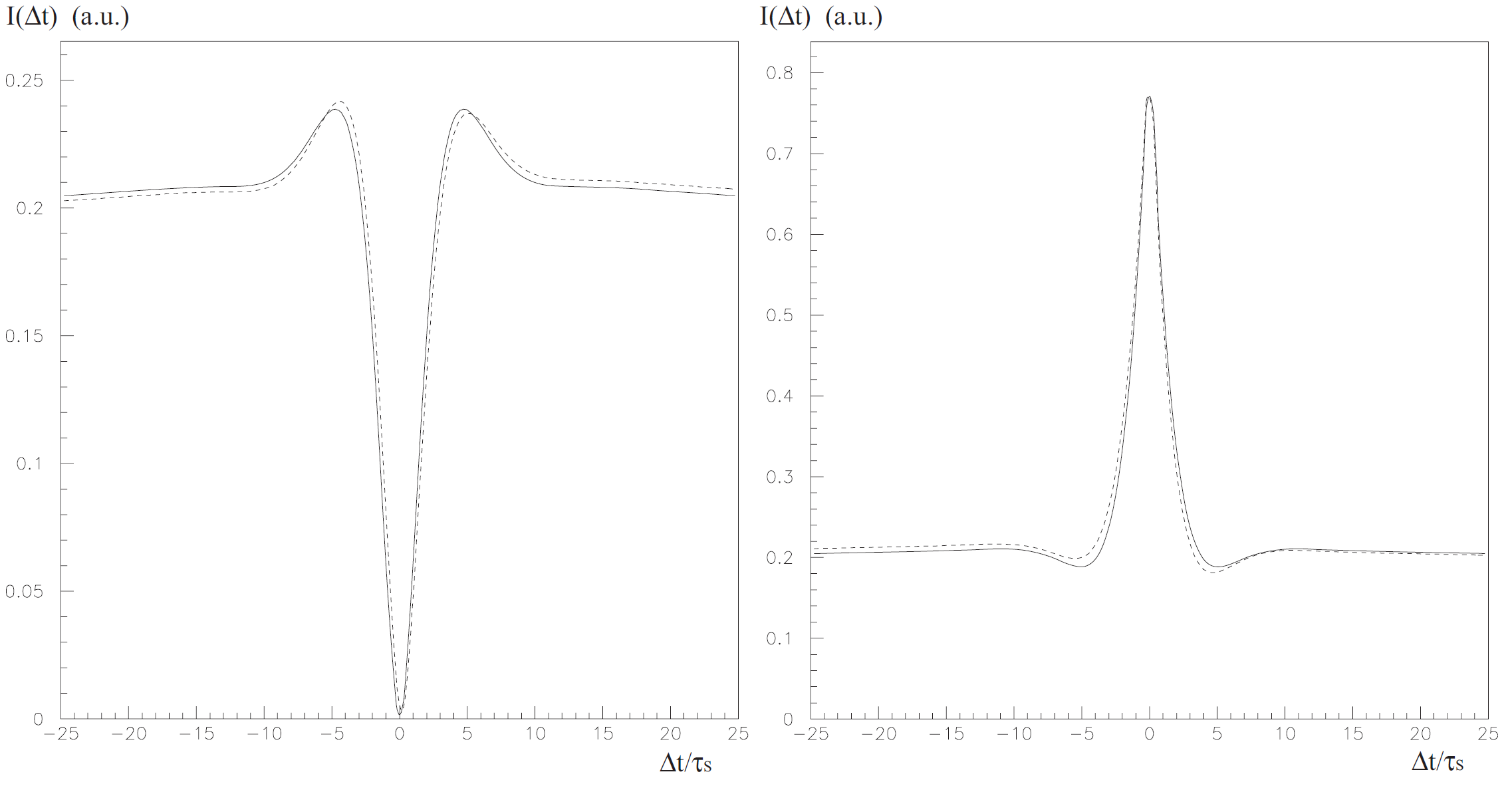}
\caption{Left: the $I(\pi^+\pi^-,\pi^0\pi^0;\Delta t$) distribution for CP-conserving events (solid line) and with small CP violation (dashed line). Right: the $I(\pi^- \ell^+ \nu,\pi^+ \ell^-\bar{\nu};\Delta t)$ distribution for CPT-conserving events (solid line) and with small CPT violation (dashed line). $\Delta t$ is given in units of $K_S$ lifetime $\tau_S$. The figures are adapted from \cite{neutral_kaon}.}
\label{fig:main_observables}
\end{figure}

If we now consider that both kaons decay into some final states $f_1$ and $f_2$ at the proper times $t_1$ and $t_2$, we can write the decay intensity distribution as a function of the decay time difference $\Delta t$ between both kaon decays \cite{neutral_kaon}:
\begin{eqnarray}
I(f_1,f_2;\Delta t)&=& \frac{C_{12}}{\Gamma_S+\Gamma_L}\Big[|\eta_1|^2 e^{-\Gamma_L \Delta t}+|\eta_2|^2 e^{-\Gamma_S \Delta t}- 2|\eta_1| |\eta_2| e^{-\frac{(\Gamma_S+\Gamma_L)}{2}\Delta t} \cos(\Delta m \Delta t+\Delta\varphi)\Big]
\label{eq:delta_t}
\end{eqnarray}
with a phase difference $\Delta\varphi=\varphi_2-\varphi_1$ and:
\begin{eqnarray}
C_{12}=\frac{|N|^2}{2}\left|\left<f_1|T|K_S\right>\left<f_2|T|K_S\right>\right|^2, \ \ \eta_i=|\eta_i|e^{i\varphi_i}\equiv\frac{\left<f_i|T|K_L\right>}{\left<f_i|T|K_S\right>}.
\label{eq:eta}
\end{eqnarray}
Eq. (\ref{eq:delta_t}) holds for $\Delta t \ge 0$, while for $\Delta t<0$ the substitutions $\Delta t\to|\Delta t|$ and $1\leftrightarrow 2$ have to be applied. Here, apart from the exponential decay terms of $K_L$ and $K_S$ we have also interference term that is characteristic at $\phi$-factories. From this distribution for various final states $f_i$ (Fig. \ref{fig:main_observables}) one can determine directly: $\Gamma_S$, $\Gamma_L$, $\Delta m$, $\eta_i$, $\Delta\varphi$ and perform tests of CP and CPT symmetries comparing experimental distributions with the theoretical predictions.

\section{Search for decoherence and CPT violation in entangled neutral kaons}

\subsection{Test of quantum coherence in $\phi\to K_L K_S\to \pi^+\pi^-\pi^+\pi^-$ decays}

\begin{figure}[!h]
\centering
\includegraphics[width=6.5cm]{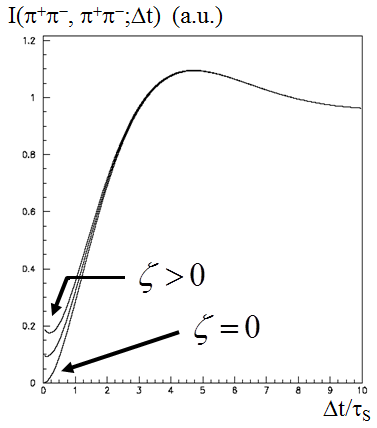}
\caption{The $I(\pi^+ \pi^-,\pi^+ \pi^-;|\Delta t|)$ distributions for quantum mechanics case, $\zeta=0$, and for two greater than zero values of the decoherence parameter. The figure is adapted from \cite{FPP6}.}
\label{double_decay_rate}
\end{figure}

If one now considers the case in which both $K_L$ and $K_S$ decay into any identical final states $f_1=f_2$, for example $K_L\to \pi^+ \pi^-$ and $K_S\to \pi^+ \pi^-$, from equation (\ref{eq:eta}) can be seen that $\eta_1=\eta_2=\eta$ and $\varphi_1=\varphi_2$. Substituting this to (\ref{eq:delta_t}) one obtains:
\begin{eqnarray}
I(f_1=f_2;|\Delta t|)= \frac{C_{12}|\eta|^2}{\Gamma_S+\Gamma_L}\Big[e^{-\Gamma_L |\Delta t|}+e^{-\Gamma_S |\Delta t|}- 2e^{-\frac{(\Gamma_S+\Gamma_L)}{2}|\Delta t|} \cos(\Delta m |\Delta t|)\Big].
\label{eq:delta_t_2}
\end{eqnarray}
The above equation implies that two kaons cannot decay into the same final states \emph{at the same time}. It is visible in Fig.~\ref{double_decay_rate}, where $I(\pi^+ \pi^-,\pi^+ \pi^-;|\Delta t|)$ is equal to 0 for $|\Delta t|=0$. What it really means is that, even though the two kaons are spatially separated, behavior of one of them is dependent on what the other does. This counterintuitive correlation is of the type first pointed out by Einstein, Podolsky and Rosen in their famous paper \cite{EPR}.

It has been suggested that the initial state soon after the $\phi$-meson decay, spontaneously factorizes to an equal weight mixture of states $\left|K_S\right> \left|K_L\right>$ or $\left|K_L\right> \left|K_S\right>$ causing decoherence. This is called the Furry's hypothesis of spontaneous factorization \cite{furry}. In general decoherence denotes the transition of a pure state into an incoherent mixture of states \cite{neutral_kaon}, meaning that entanglement of particles is lost. The decoherence parameter $\zeta$ can be introduced by multiplying the interference term by a factor $(1-\zeta)$. In general $\zeta$ depends on the basis in which the initial state is expressed: $\{\left|K^0\right>, \left|\bar{K}^0\right>\}$ or $\{\left|K_S \right>, \left|K_L\right>\}$, and in $K_S K_L$ basis it reads \cite{neutral_kaon}:
\begin{eqnarray}
I(\pi^+\pi^-,\pi^+\pi^-;\Delta t)\propto e^{-\Gamma_L \Delta t}+e^{-\Gamma_S \Delta t}- 2(1-\zeta_{SL})e^{-\frac{(\Gamma_S+\Gamma_L)}{2}\Delta t} \cos(\Delta m \Delta t).
\label{eq:decoherence}
\end{eqnarray}
A value of $\zeta=0$ corresponds to the usual quantum mechanics case, $\zeta=1$ to the total decoherence and different values to intermediate situations between these two. Hence, the decoherence parameter measures the amount of deviation from the predictions of quantum mechanics. Fig.~\ref{double_decay_rate} shows sensitivity of the double decay rate distribution to the value of $\zeta$. The biggest discrepancy is for $\Delta t$ close to 0.

\begin{figure}[!h]
\centering
\includegraphics[width=6.5cm]{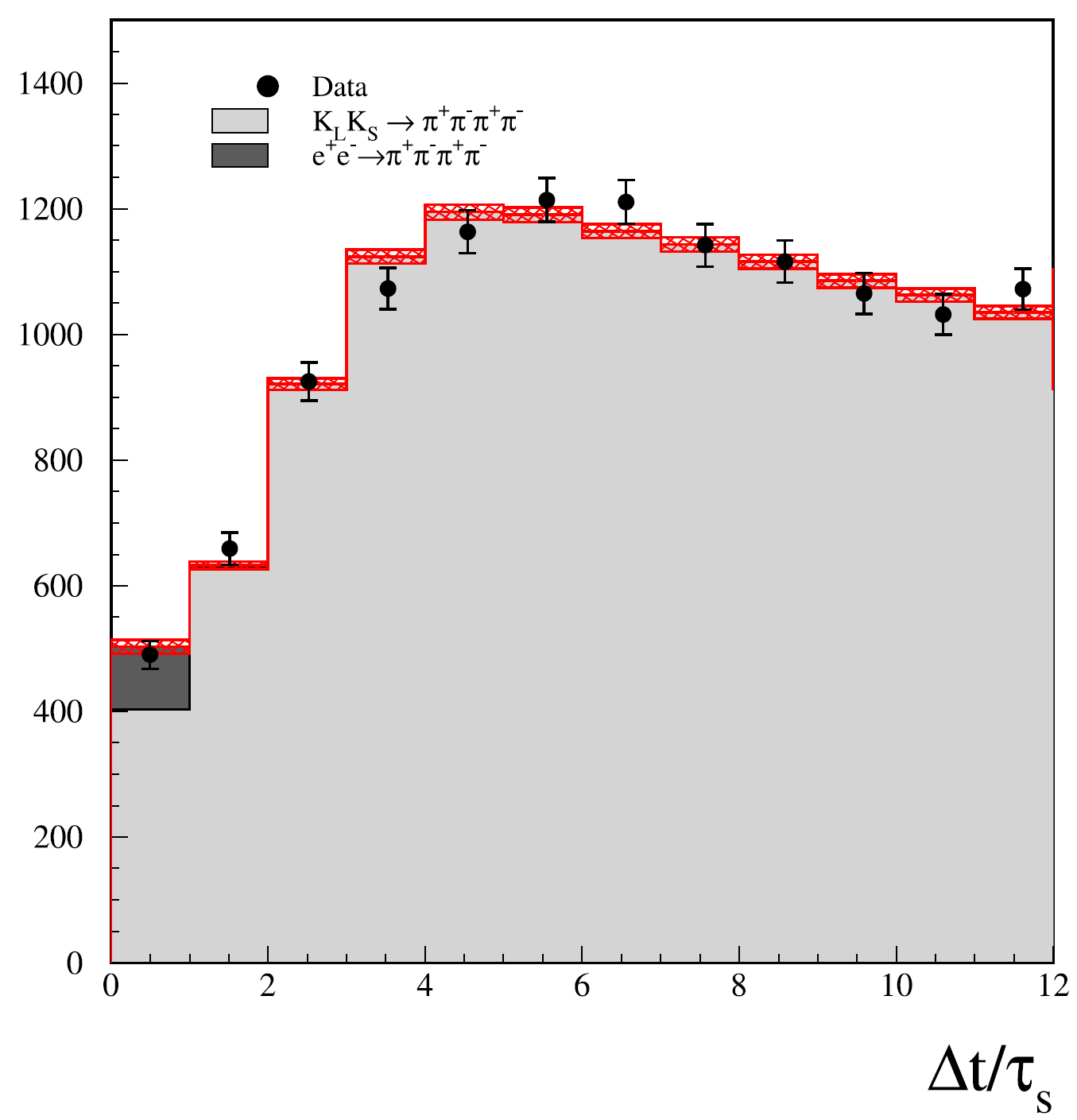}
\caption{Fit to $\Delta t$ distribution of the events $\phi\to K_L K_S\to \pi^+\pi^-\pi^+\pi^-$. Points denote experimental results and histogram shows results of the MonteCarlo simulations. This binning was chosen due to the time resolution $\sigma(\Delta t)\sim\tau_S$. The figure is adapted from \cite{first_obs}.}
\label{fig:decoh_data}
\end{figure}

At KLOE tests of the coherence were performed by analyzing data corresponding to the $\sim$1.5~fb$^{-1}$ of integrated luminosity. The determined experimental distribution of the $\phi\to K_L K_S\to \pi^+\pi^-\pi^+\pi^-$ intensity as a function of the absolute value of $\Delta t$ is shown in Fig.~\ref{fig:decoh_data}. Current measurements show that there are no deviations from quantum mechanics \cite{CPT_and_GM_tests}:
\begin{eqnarray}
\zeta_{SL}&=&\left(0.3\pm1.8_{stat}\pm 0.6_{syst}\right)\cdot 10^{-2},\nonumber \\
\zeta_{0\bar{0}}&=&\left(1.4\pm 9.5_{stat}\pm 3.8_{syst}\right)\cdot 10^{-7}.
\end{eqnarray}
This result can be compared to the one obtained by the CPLEAR data \cite{PR}: $\zeta_{0\bar{0}}=0.4\pm0.7$ and the BELLE collaboration result measured in the B meson system \cite{PRL}: $\zeta_{0\bar{0}}^B=0.029\pm0.057$.

\subsection{CPT violation in entangled K states}

There are several hypothesis of possible CPT violation sources, one of them related to quantum gravity effects that could induce loss of information about the initial state, that is in striking conflict with quantum mechanics and its unitarity principle. This decoherence necessarily implies CPT violation in the sense that the quantum mechanical operator generating CPT transformation cannot be consistently defined. The resulting loss of particle-antiparticle identity could induce a breakdown of the correlation of initial state imposed by Bose statistics. As a result the initial state gains small symmetric term \cite{Bernabeu}:
\begin{eqnarray}
\left|i\right>&\propto&(K^0\bar{K}^0-\bar{K}^0 K^0)+\omega(K^0\bar{K}^0+\bar{K}^0 K^0) \nonumber \\
&\propto&(K_S K_L-K_L K_S)+\omega(K_S K_S-K_L K_L)
\end{eqnarray}

\begin{figure}[!h]
\centering
\includegraphics[width=12cm]{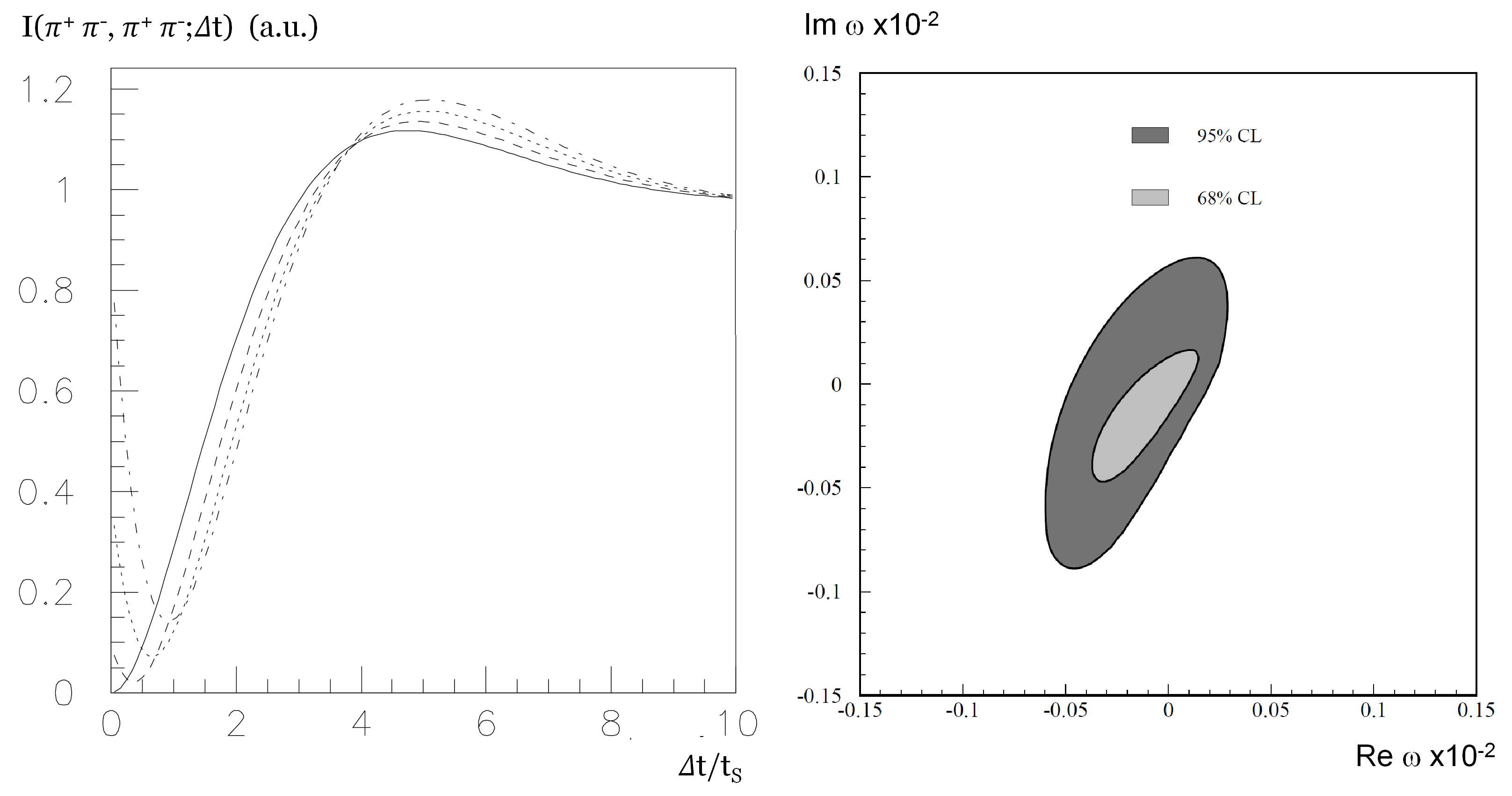}
\caption{Left: the theoretical $I(\pi^+ \pi^-,\pi^+ \pi^-;\Delta t,\omega)$ distribution without CPT violation (solid line), $\omega=0$, and for three different than zero values of $\omega$ (dashed lines) is shown (the figure is adapted from \cite{FPP6}). Right: a contour plot of $\Im \omega$ versus $\Re\omega$ at the 68\% and 95\% of confidence level (the figure is adapted from \cite{CPT_and_GM_tests}).}
\label{fig:omega}
\end{figure}

The parameter $\omega$ is a new complex CPT violation parameter that could be measured only in entangled systems. One expects that it is at most \cite{neutral_kaon}: $|\omega|^2=O\left(\frac{E^2 / M_{Planck}}{\Delta\Gamma}\right)\approx 10^{-5} \Rightarrow |\omega|\sim 10^{-3}$. In some specific microscopic models of space-time foam arising from non-critical string theory it is predicted to be up to \cite{Mavromatos}: $|\omega|\approx10^{-4}-10^{-5}$. For the omega parameter the maximum sensitivity is expected, as for the decoherence parameter, for $K_S K_L\to\pi^+\pi^-\pi^+\pi^-$ decays (Fig. \ref{fig:omega} left).

This analysis was performed by the KLOE collaboration on the same $I(\pi^+\pi^-\pi^+\pi^-;\Delta t)$ distribution as before, using $\sim$1.5~fb$^{-1}$ of data, by fitting the decay intensity distribution modified including $\omega$ parameter. The obtained result (Fig. \ref{fig:omega} right) is consistent with no CPT violation effects \cite{CPT_and_GM_tests}:
\begin{eqnarray}
\Re \omega&=&\left(-1.6^{+3.0}_{-2.1 \ STAT}\pm 0.4_{SYST}\right)\cdot 10^{-4}, \nonumber\\
\Im \omega&=&\left(-1.7^{+3.3}_{-3.0 \ STAT}\pm 1.2_{SYST}\right)\cdot 10^{-4}. 
\end{eqnarray}
The upper limit at 95\% confidence level for the module is $|\omega|\le1.0 \cdot 10^{-3}$. The accuracy already reaches the interesting Planck scale region. In comparision, in the B meson system only the real part of it was estimated and with minor precision \cite{Nebot}: $-0.0084\le\Re \omega\le 0.0100$.

\section{$K_S$ regeneration at KLOE}
KLOE results on decoherence and CPTV parameters are dominated by statistical errors. At KLOE-2 this error is expected to be reduced by about factor 10 and systematical uncertainties will play the dominant role. One of the main sources of systematic uncertainty is the $K_S$ regeneration. It affects measurement of the CPV channel, $K_L\to\pi^+\pi^-$, in the sense that when $K_L$ reaches the regenerating material, it can regenerate into $K_S$ and then almost immediately decay into $K_S^{reg}\to\pi^+\pi^-$. These decays of regenerated $K_S$ can be confused as CPV decays of $K_L$. This background have to be carefully rejected and subtracted from the measured distribution (Fig.~\ref{fig:decoh_data}).

\begin{figure}[!h]
\centering
\includegraphics[width=13cm]{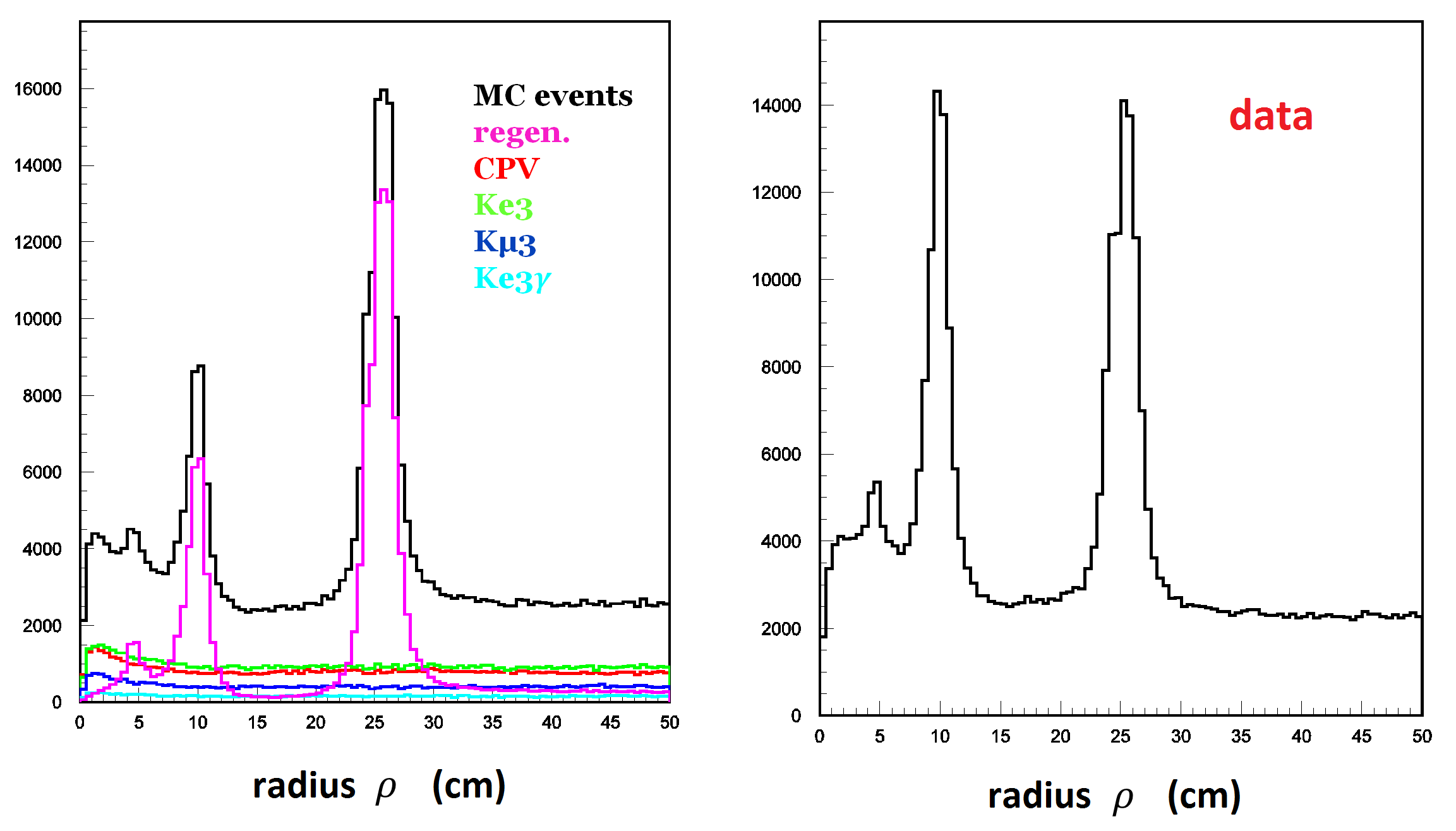}
\caption{The transverse radius distribution for registered $\phi\to K_L K_S\to \pi^+\pi^-\pi^+\pi^-$ decays from the MonteCarlo simulations (left) and data (right). For the cylindrical elements, so beryllium beam pipe and inner wall of the drift chamber, the regenerated events appear as symmetrical peaks and are situated at 4.4~cm and 25~cm, respectively. Spherical beam pipe corresponds to the peak at 10~cm. The figure is adapted from \cite{thesis}.}
\label{fig:rho}
\end{figure}

There are three main regenerators at KLOE: the cylindrical and the spherical beam pipes, described before, and the inner wall of the drift chamber. On the transverse radius distribution (Fig.~\ref{fig:rho}), $\rho=\sqrt{x^2+y^2}$, obtained with loose selection cuts, the regenerated events appear as peaks on the exponential background. From the analysis of these distributions one can evaluate the regeneration cross section and the impact of regeneration on the decoherence and CPTV parameters.

\section{KLOE-2 plans}

The upgraded KLOE detector, KLOE-2, at the DAFNE machine upgraded in luminosity, is about to start taking data. The physics program \cite{EPJC} is extended in comparison to the KLOE one. The DAFNE accelerator is assumed to deliver an integrated luminosity up to about 20~fb$^{-1}$ during the next 3 years. Thanks to a new inner tracker detector \cite{tech_design} there will be possible to improve resolution of the decay vertex reconstruction by about 3 times. This, in turn, will allow to improve the resolution on $\Delta t$ and consequently the sensitivity to parameters of the interference.

\section{Conclusions}

The entangled neutral kaon system is an excellent laboratory for the study of CPT symmetry and the basic principles of quantum mechanics. Several parameters related to possible decoherence and CPT violation (due to quantum gravity effects) have been measured at KLOE, in same cases with a precision reaching the interesting Planck scale region. All results are consistent with no CPT violation and no decoherence.

KLOE-2 at DAFNE upgraded in luminosity is going to start taking data. Neutral kaon interferometry, CPT symmetry and quantum mechanics tests are one of the main issues of the KLOE-2 physics program.\\

\textbf{Acknowledgments}\\
We acknowledge support by Polish Ministry of Science and Higher Education through the Grant No. 0469/B/H03/2009/37, by MPD program of Polish Science Foundation and by the European Community-Research Infrastructure Integrating Activity ''Study of Strongly Interacting Matter'' (acronym HadronPhysics2, Grant Agreement n. 227431) under the Seventh Framework Programme of EU.

\end{document}